# Soft Phonon Mode Dynamics in Aurivillius Type Structures

Deepam Maurya[1], Ali Charkhesht[2], Sanjeev K. Nayak[3], Fu-Chang Sun[3], Deepu George[2], Abhijit Pramanick[4], Min-Gyu Kang[1], Hyun–Cheol Song[1], Marshall M. Alexander[2], Djamila Lou[2], Giti A. Khodaparast[2], S. Pamir Alpay[3*], N. Q. Vinh[2*], and Shashank Priya[1*]

[1]*Bio-inspired Materials and Devices Laboratory (BMDL), Center for Energy Harvesting Materials and Systems (CEHMS), Virginia Tech, Blacksburg, VA 24061 USA*
[2]*Department of Physics, Virginia Tech, Blacksburg, VA 24061 USA*
[3]*Department of Materials Science & Engineering, Department of Physics, Institute of Materials Science, University of Connecticut, Storrs, CT 06269-3136, USA*
[4]*Department of Physics and Materials Science, City University of Hong Kong, Kowloon, Hong Kong SAR*

We report the dynamics of soft phonon modes and their role towards the various structural transformations in Aurivillius materials by employing terahertz frequency-domain spectroscopy, atomic pair distribution function analysis, and first-principles calculations. We have chosen $Bi_4Ti_3O_{12}$ as a model system and identified soft phonon modes associated with the paraelectric tetragonal to the ferroelectric monoclinic transition. Three soft phonon modes have been discovered which exhibit a strong temperature dependence. We have determined that the anharmonicity in Bi−O bonds plays a significant role in phonon softening and that Bi cations play an important role in the emergence of ferroelectricity.

The knowledge of soft phonon mode properties is crucial for understanding the origin of lattice instabilities and structural phase transitions in bismuth layered ferroelectrics (Aurivillius-type structures represented as $[Bi_2O_2][A_{m-1}B_mO_{3m+1}]$, where m = 3, A = Bi, and B=Ti, for $Bi_4Ti_3O_{12}$). Typically, ferroelectric-paraelectric phase transitions in these materials occur with the heavily damped phonons in the terahertz (THz) frequencies [1,2]. Additionally, there could be subtle structural distortions below Curie temperature ($T_c$), which are often difficult to correlate with phonon dynamics. Since structural changes drive many material properties, a fundamental understanding of dynamics of these phonon modes is critical for designing high performance ferroelectric materials and devices [3]. The number of phonon modes are defined by the nature of changes in the symmetry during the transitions. The phase transitions involving more than one soft phonon modes [4] and corresponding order parameters, may induce structural transformations at temperatures below $T_c$. However, the condensation of more than one phonon modes at a single transition is quite unusual [5].

Here, we employed $Bi_4Ti_3O_{12}$ (BiT) as a model system to understand phonon modes related to phase transitions in Aurivillius materials. Interest in layered structures, such as Aurivillius compounds, Dion-Jacobson phases, and Ruddlesden-Popper phases, has been increasing rapidly due to their relevance for 2D materials based electronics along with high Curie temperature ($T_c$ for BiT > 600º C) [6]. The ferroelectric members in these families have potential for high temperature sensors and fatigue-free ferroelectric memory devices etc. [7]. Furthermore, these layered materials exhibit anisotropic and very low thermal conductivity due to effective phonon scattering [8]. The structure of BiT consists of perovskite-like block $(Bi_2Ti_3O_{10})^{2-}$ interleaved with fluorite like $(Bi_2O_2)^{2+}$ layers perpendicular to pseudo-tetragonal $c$-axis [9] which results in relatively higher polarization [10]. In terms of phase transformation characteristics, BiT undergoes a ferroelectric phase transformation from the high temperature tetragonal paraelectric phase to a lower temperature polar phase [11]. This phase transition involves displacement of Bi atoms within the perovskite layers and the rotation of the $TiO_6$ octahedra. [12]

Using a sensitive and high resolution THz frequency-domain spectroscopy, we have experimentally discovered the so far elusive three phonon modes in BiT system. These phonon modes, not reported earlier, are expected to have important implications towards the symmetry breaking from the high temperature tetragonal to the low temperature monoclinic phase as well as structural transformations below $T_c$. We have further employed atomic pair distribution function (PDF) analysis to correlate the dynamics of Bi ions with the observed phonon dynamics. These results are complemented by first principles based phonon studies, which describe the THz spectroscopy and identify that the main contribution to the atom-projected phonon density of states (DOS) comes from the Bi atoms.

Much effort has been devoted to understand structural changes with respect to temperature in Aurivillius ferroelectric materials. Theoretical [5] and experimental [13] studies have reported possible triggered phase transitions from low temperature polar monoclinic phase to high temperature tetragonal phase of BiT. The low temperature ferroelectric monoclinic phase of BiT requires condensation of at least three different symmetry breaking modes, which have hitherto not been observed experimentally [5]. An observation of the temperature dependence of the lowest frequency polar phonon mode (denoted as a soft phonon mode), using Raman scattering is not very convincing, because, the intensity of the soft phonon mode decreases rapidly with increasing temperature [4]. On the other hand, THz frequency-domain technique, used in this study provides direct measurement of optical phonons providing opportunity to understand the basic nature of transformations.

Prior studies have investigated the dynamics of the ferroelectric transition in Bi-layered ferroelectric materials using THz time-domain spectroscopy [4,14], where only one optical soft mode was observed in the ferroelectric phase of the BiT material [4], which was underdamped above the phase transition temperature ($T_c$) due to the change of selection rules in the paraelectric phase. However, using our high resolution and large dynamic range THz frequency-domain spectroscopy [15], we have observed multiple optical modes which could explain the various structural transformations leading to the ferroelectric phase in BiT and BiT-like layered materials. We further employed atomic pair distribution function (PDF) analysis and the first-principles calculations to provide the fundamental understanding of phonon dynamics in layered ferroelectrics.

The THz experiments were performed on highly textured (00l) oriented BiT ceramics (see supplemental material (SM) for details). To confirm phase formation and texture, the X-ray powder diffraction (XRD) spectra were recorded at room temperature by using a Philips Xpert Pro x-ray diffractometer (Almelo, The Netherlands), as shown in Fig. S-1. For PDF analysis, high resolution powder x-ray diffraction data was recorded using beamline 11-BM at Argonne National Laboratory. The surface morphology of sintered samples was observed using a LEO Zeiss 1550 (Zeiss, Munich, Germany) scanning electron microscope. Transmission electron microscopy was performed using the FEI Titan 300 electron microscope. The THz frequency-domain spectrometer employs a commercial Vector Network Analyzer from Agilent, the N5225A PNA which covered the frequency range from 10 MHz to 50 GHz, and THz frequency extenders as well as matched harmonic detectors developed by Virginia Diodes, Inc. with frequency range from 60 GHz to 1.12 THz. The dynamic range of the instrument reaches $10^{13}$ with a spectral resolution of less than 100 Hz (see SM).

Figure 1a shows a bright field cross-section TEM image of textured BiT samples. The cross-section morphology indicates that the plate-type BiT grains are stacked along the thickness of the sample confirming the textured microstructure of the BiT. From these images the size of the plate-type grains is in the range of 5−15 μm (Fig. S-2 in SM). The thickness of these plate type grains was found to be in the range of 200–500 nm (Fig. 1a). The stacking of the pseudo-perovskite and $(Bi_2O_2)^{2+}$ layers was clearly observed in the HR-TEM lattice fringes from [100] zone axis, as shown in Fig. 1b. The upper inset in Fig. 1b shows the FFT pattern indicating [100] zone axis, whereas, the lower inset indicates a TEM image with low magnification, revealing layered structure. Due to the two-fold in-plane symmetry, the distinctive stacking of the pseudo-perovskite and $(Bi_2O_2)^{2+}$ layers was not observed (Fig. S-2 in SM) from [001] zone axis. The schematic of BiT layered structure at low temperature phase is provided in Fig. 1c.

The high dynamic range and high resolution of our THz frequency-domain spectroscopy allows us to observe the lowest-frequency polar phonon modes or soft phonon modes. The refractive index, $n(v)$, and absorption coefficient, $\alpha(v)$, of BiT samples have been determined through THz measurements, as shown in Figs 2a and 2b, respectively for several temperatures from room temperature to near $T_c$ at 600ºC. From the absorption and refractive index measurements, we have defined the complex dielectric response of the sample. The frequency-dependent complex dielectric response, $\varepsilon^*(v) = \varepsilon'(v) - i\varepsilon''(v)$, is related to the complex refractive index, $n^*(v) = n(v) - i\kappa(v)$, through the relations:

$$\begin{aligned} \epsilon'_{sol}(v) &= n^2(v) - \kappa^2(v) = n^2(v) - (c\alpha(v)/4\pi v)^2 \\ \epsilon''_{sol}(v) &= 2n(v) \cdot \kappa(v) = 2n(v)c\alpha(v)/4\pi v \end{aligned} \tag{1}$$

where $v$ is the frequency of the THz radiation. The real part, $n(v)$, is the refractive index and the imaginary part, $\kappa(v)$, is the extinction coefficient and indicates the attenuation when the radiation propagates through the material. The extinction coefficient, $\kappa(\nu)$, is related to the absorption coefficient through a relation: $\alpha(\nu) = \frac{4\pi \cdot \nu \cdot \kappa(\nu)}{c}$, where $c$ is the speed of light.

Accounting these relationships, we have obtained the complex dielectric response of the BiT sample including the dielectric loss, $\varepsilon''(\nu)$ and permittivity $\varepsilon'(\nu)$ as a function of THz frequency at various temperatures up to Curie temperature of the material (Figs 2c and 2d), respectively. Unlike the previous reports where only one mode was reported at 0.83 THz [4,7], we have observed three phonon modes at 0.68 THz (22.68 cm$^{-1}$), 0.86 THz (28.69 cm$^{-1}$) and 0.96 THz (32.02 cm$^{-1}$) at room temperature for the 30 μm BiT sample. The observations were reproducible (3 different samples) and the temperature cycling did not noticeably affect the observed phonon modes. A strong temperature dependence of these phonon modes and the corresponding phonon mode frequencies decreasing toward zero near the phase transition temperature, $T_c$, appear to suggest their soft mode behavior. The theoretical calculations further confirm soft nature of the phonon modes. Upon heating, in addition to mode shifting, the full width at half maximum (FWHM) of the absorption peak also increases with temperature.

To gain better insight into the damping process of the three soft phonon modes, we have fitted the complex dielectric response obtained from our THz frequency-domain spectroscopy at various temperatures. For this, we employed a function containing a sum of three damped Lorentz oscillators describing the optical phonons of the ferroelectric materials [16]:

$$\varepsilon^*(\nu) = \varepsilon_\infty + \sum_{j=1}^{3} \frac{A_j/(2\pi)^2}{\nu_j^2 - \nu^2 + i\nu(\gamma_j)} \qquad (2)$$

where $A_j/(2\pi)^2$, $\nu_j$, and $\gamma_j$ are, respectively, the spectral amplitude of the $j$ damped resonances, its frequency, and its damping coefficient, $\varepsilon_\infty$ describes contributions to the dielectric function from modes at frequencies much greater than our experimental range. The parameters of three soft-phonon modes as function of sample temperature are summarized in Figs 2*e*, 2*f*, and 2*g*. The resonant frequencies of the three soft phonon modes ($\nu_1$, $\nu_2$, and $\nu_3$) decreases with increasing sample temperature. The damping of these modes increases with the sample temperature. While the damping and FWHM values for the $\nu_1$ phonon mode slightly changes with the sample temperature, these parameters for the $\nu_2$, and $\nu_3$ modes exhibit a strong increase with increasing temperature, which suggests a finite coupling between these modes. The results provide an evidence of the soft nature of these phonon modes.

The pronounced softening of the $\nu_2$ and $\nu_3$ modes with increasing temperature can be understood as a result of impending ferroelectric-to-paraelectric phase transition as the sample temperature approaches $T_c \sim 700$ °C. The anomalies below $T_c$ could be observed from the THz spectra, as shown in Fig. 2. A discontinuity in the temperature dependence and a sharp increase in FWHM for all these modes occur at T > 300 °C. The damping of these phonon modes increase significantly when the sample temperature reaches near to $T_c$. In order to understand the possible structural distortions, which might explain anomalies observed below $T_c$, atomic PDF analysis was performed. The PDF measurements obtained from a total scattering XRD pattern via a Fourier transform provides us an approach to study the local structure of materials. Because the total scattering pattern is composed of Bragg as well as diffuse scattering contributions, the information contains local, medium range and long range structure information. The high energy XRD results were corrected for the sample absorption, background, Compton scattering, and incident flux. The intensities were normalized and reduced to the structure factor S(Q) (where Q is the diffraction wave vector), which was Fourier transformed to the corresponding PDFs using

PDFgetX[17], G(r) (Fig. S-3 in SM). The G(r) gives the probability of finding a pair of atoms at a distance r [18]:

$$G(r) = \frac{2}{\pi} \int_0^\infty Q[S(Q) - 1] \sin(Qr)\, dQ \qquad (3)$$

Having the experimental PDF, one usually wants to determine local structural changes. The PDF results were fitted with B2cb structure having lattice parameters a = 5.448 Å, b = 5.411 Å, c = 32.83 Å, as shown in Fig. 3a. The atomic positions were same as given by Rae et al. [19]. The peaks for the nearest neighbors are highlighted in Fig. 3b. The inset of Fig. 3b shows the experimental PDF, G(r), for the BiT at different conditions. One can clearly see the broadening of the peaks related to the bismuth oxide layer and perovskite layer at 400 °C (Fig. 3b). The broadening of these peaks indicates increasing disorder in bismuth layers. Most notably, the peak related to the perovskite layer is not just broadened, but also, became asymmetric indicating increased anharmonicity of the Bi−O bonds. We have determined that the anharmonicity of these bonds plays a significant role in shifting of the soft phonon modes, and could possibly be the origin of anomalies observed in THz spectra below $T_c$.

In order to obtain further insights into the experimentally observed phonon dynamics and bond anharmonicity, phonon studies were performed with first principles density functional theory (DFT) [20,21]. The generalized gradient approximation (GGA) [22] was used as the exchange-correlation functional together with the projector-augmented wave method [23] as implemented in the Vienna *Ab initio* Simulation Package (VASP) [24-26]. The primitive cell dimensions for the monoclinic low temperature phase of BiT with space group *P*c were found to be *a* = 5.49 Å, *b* = 5.53 Å, *c* = 16.88 Å, and $\alpha_0$ = 80.61°. These values are in good agreement with experimental reports [27] and other first-principles computations [28]. Phonon calculations were performed by linear response method [29] and the frozen phonon method together with Phonopy [30] (see SM). Combination of DFT with frozen phonon method provides the platform to analyze lattice dynamics in quasi harmonic approximation with the inter-atomic forces calculated from the state-of-the-art electronic structure methods.

The phonon density of states (DOS) is shown as a black solid line in Fig. 4 (also see SM). We identify three peaks P1, P2 and P3 in the range 0−1.5 THz. An analysis from the band structure suggests that it is plausible to match P1 to the optical mode at the Z point of the Brillouin zone (BZ) while P2 can be matched with two phonon modes at the Γ point of the BZ which are almost degenerate (Fig. S-5 in SM). P3 is an optical mode at the Γ point. The phonon eigenvalues at the Γ point from density functional perturbation theory (DFPT) calculations for a 2×2×1 supercell is found to be 1.11 THz, 1.13 THz and 1.26 THz for the three low energy phonon modes. The symmetry of the three modes at the Γ point is found to be $A''$, $A'$, and $A'$, all of which are IR active. In order to observe the effect of the volume change, we computed the phonon DOS with a volume change of +1.5 % $V_0$ and −1.5 % $V_0$ shown as red and green solid lines (Fig.4), respectively. The octahedral tilting plays a significant role in the phase transition of the BIT system, (see SM). We have also explored the effect of octahedral tilting on the phonon DOS by increasing the monoclinic angle $\alpha_0$ by 0.4% and 0.7%, as shown in Fig. 4 in blue and magenta solid lines, respectively. These theoretical calculations suggest that the phonon peaks are shifted to lower frequencies for the models deviating from the ground state monoclinic structure. The peak P2 appears to split the constituting phonon modes by about 0.2 THz when the volume is increased due to the atomic rearrangement in relaxed lattice. On reduction of the volume to 0.985 $V_0$, the phonons were found to exhibit hardening behavior. The analysis of energy as a function of mode amplitude for the three low energy phonon modes, for the models with increased lattice angle with respect to the ground state monoclinic value ($\alpha_0$), suggested two soft phonons (Fig. S-6 in SM). These results indicate soft phonons in BiT which appear with changes in both volume and the octahedral tilting. The atomic rearrangement accommodating these changes could lead to the anharmonicity in the interatomic bonds, as observed in the PDF analysis of Fig. 3.

The phonon modes for P2 and P3 of Fig. 4 are shown in SM (animated gif files) where the vibration of ions in the perovskite- and fluorite-blocks are displayed. The correlated motion of atoms within each block, which are out-of-phase among each other, is shown in the gray and green regions in Fig. 1c. The out-of-phase oscillations of the lattice blocks could potentially lead to the deviation from the monoclinic towards tetragonal phase (Fig. 1d). The structural change could be described using the lattice parameter transformation $a < b \rightarrow a' = b'$, such that the lattice parameter of the tetragonal phase is $a'$ ($=b'$) = $a/\sqrt{2}$ (Fig. 1e). This mechanism is consistent with the theoretical findings of Ref. [31], where it is suggested that two unstable $E_u$ modes in BiT, one involving the motion of fluorite layers in a direction relative to the perovskite $(TiO_6)^{8-}$ blocks and the second mode involving the motion of the Bi ions in the perovskite A site with respect to the perovskite blocks, are responsible for the phase change. The atomic displacements in the fluorite layers are larger than in the perovskite layers for the three calculated modes in our study. Thus, we underline that the chemical nature of large cation in the fluorite layers in the Aurivillius family and similar layered oxides is crucial for structural transformations. We note, however, that the theoretical tools used here could have certain limitations while applying for more complicated structures. Firstly, the quasi-harmonic approximation is not appropriate for larger scale volume and/or angular variations. In addition, this approach may not be applicable to accurately determine phase transformation temperatures in strongly correlated systems. Future improvements on the current theoretical foundations will therefore be necessary to tackle more complex systems.

In summary, we have probed dynamics of soft phonon modes and its role in the structural transformations on highly textured (00l) oriented $Bi_4Ti_3O_{12}$ using the THz frequency-domain spectroscopy. The results from the THz frequency-domain spectroscopy have revealed three low frequency soft phonon modes, which have been supported from first-principles study. The anharmonicity of the Bi−O bonds plays a leading role in these low frequency phonon modes with majority of contribution to the phonon density of states comes from the Bi atoms. The fundamental understanding about various factors affecting phonon dynamics and structural changes described here provides useful information in designing tailored phase transition and functionality (e.g. ferroelectric and thermal properties) of layered-structure ferroelectric materials.

**Acknowledgments:** This work was supported by the AFOSR through grant FA9550-14-1-0376. The terahertz-dielectric study was supported by the Institute of Critical Technology and Applied Sciences (ICTAS) at Virginia Tech. MGK and HCS acknowledge financial support through the Department of Energy program (DE-FG02-06ER46290). The computational resources from the Taylor L. Booth Engineering Center for Advanced Technology (BECAT) at University of Connecticut is gratefully acknowledged. One of the authors (F.-C. S) would like to thank J. Skelton at University of Bath and H. Tran and K. Pitike at University of Connecticut for helpful discussions. S.K.N would like to acknowledge technical support from Serge M. Nakhmanson, University of Connecticut and Waheed A. Adeagbo, Martin Luther University Halle-Wittenberg. Use of the Advanced Photon Source at Argonne National Laboratory was supported by the U. S. Department of Energy, Office of Science, Office of Basic Energy Sciences, under Contract No. DE-AC02-06CH11357. AP gratefully acknowledges funding support from CityU Start-up Grant for New Faculty (Project Number 7200514). A. C. and S.K.N. contributed equally to this work. S.P. acknowledges the support from NSF CREST program.

## References

[1]     J. Hlinka, T. Ostapchuk, D. Nuzhnyy, J. Petzelt, P. Kuzel, C. Kadlec, P. Vanek, I. Ponomareva, and L. Bellaiche, Physical Review Letters **101**, 167402 (2008).

[2]     D. Wang, A. A. Bokov, Z. G. Ye, J. Hlinka, and L. Bellaiche, Nat Commun **7** (2016).

[3]     M. S. Senn, D. A. Keen, T. C. A. Lucas, J. A. Hriljac, and A. L. Goodwin, Physical Review Letters **116**, 207602 (2016).

[4]     D. Nuzhnyy *et al.*, Phys Rev B **74** (2006).

[5]     J. M. Perez-Mato, P. Blaha, K. Schwarz, M. Aroyo, D. Orobengoa, I. Etxebarria, and A. García, Phys Rev B **77**, 184104 (2008).

[6]     W. S. Choi and H. N. Lee, Phys Rev B **91**, 174101 (2015).

[7]     B. H. Park, B. S. Kang, S. D. Bu, T. W. Noh, J. Lee, and W. Jo, Nature **401**, 682 (1999).

[8]     C. Chiritescu, D. G. Cahill, N. Nguyen, D. Johnson, A. Bodapati, P. Keblinski, and P. Zschack, Science **315**, 351 (2007).

[9]     E. J. Nichols, J. W. J. Shi, A. Huq, S. C. Vogel, and S. T. Misture, J Solid State Chem **197**, 475 (2013).

[10]    J. H. Lee, R. H. Shin, and W. Jo, Phys Rev B **84**, 094112 (2011).

[11]    A. Shrinagar, A. Garg, R. Prasad, and S. Auluck, Acta Crystallogr A **64**, 368 (2008).

[12]    Qingdi Zhou and Brendan J. Kennedy, Chem. Mater. **15,** 4025 (2003).

[13]    M. Iwata, K. Ando, M. Maeda, and Y. Ishibashi, Journal of the Physical Society of Japan **82**, 025001 (2013).

[14]    M. Kempa, P. Kuzel, S. Kamba, P. Samoukhina, J. Petzelt, A. Garg, and Z. H. Barber, J Phys-Condens Mat **15**, 8095 (2003).

[15]    D. K. George, A. Charkhesht, and N. Q. Vinh, Review of Scientific Instruments **86**, 123105 (2015).

[16]    N. Q. Vinh, M. S. Sherwin, S. J. Allen, D. K. George, A. J. Rahmani, and K. W. Plaxco, The Journal of Chemical Physics **142**, 164502 (2015).

[17]    P. Juhas, T. Davis, C. L. Farrow, and S. J. L. Billinge, Journal of Applied Crystallography **46**, 560 (2013).

[18]    Y. Yasuhiro, K. Shinji, and M. Jun'ichiro, Japanese Journal of Applied Physics **45**, 7556 (2006).

[19]    A. D. Rae, J. G. Thompson, R. L. Withers, and A. C. Willis, Acta Crystallographica Section B **46**, 474 (1990).

[20]    W. Kohn and L. J. Sham, Physical Review **140**, A1133 (1965).

[21]    P. Hohenberg and W. Kohn, Physical Review **136**, B864 (1964).

[22]    J. P. Perdew, K. Burke, and M. Ernzerhof, Physical Review Letters **77**, 3865 (1996).

[23]    P. E. Blöchl, Phys. Rev. B **50**, 17953 (1994).

[24]    G. Kresse and D. Joubert, Phys Rev B **59**, 1758 (1999).

[25]    G. Kresse and J. Furthmüller, Computational Materials Science **6**, 15 (1996).

[26]    G. Kresse and J. Furthmüller, Phys Rev B **54**, 11169 (1996).

[27]    J. Min Ku, K. Yong-Il, N. Seung-Hoon, S. Jung Min, J. Chang Hwa, and W. Seong Ihl, Journal of Physics D: Applied Physics **40**, 4647 (2007).

[28]    D. J. Singh, S. S. A. Seo, and H. N. Lee, Phys Rev B **82**, 180103 (2010).

[29]    S. Baroni and R. Resta, Phys Rev B **33**, 7017 (1986).

[30]    A. Togo and I. Tanaka, Scripta Materialia **108**, 1 (2015).

[31]    R. Machado, M. G. Stachiotti, R. L. Migoni, and A. H. Tera, Phys Rev B **70**, 214112 (2004).

**Figures and Figure Captions**

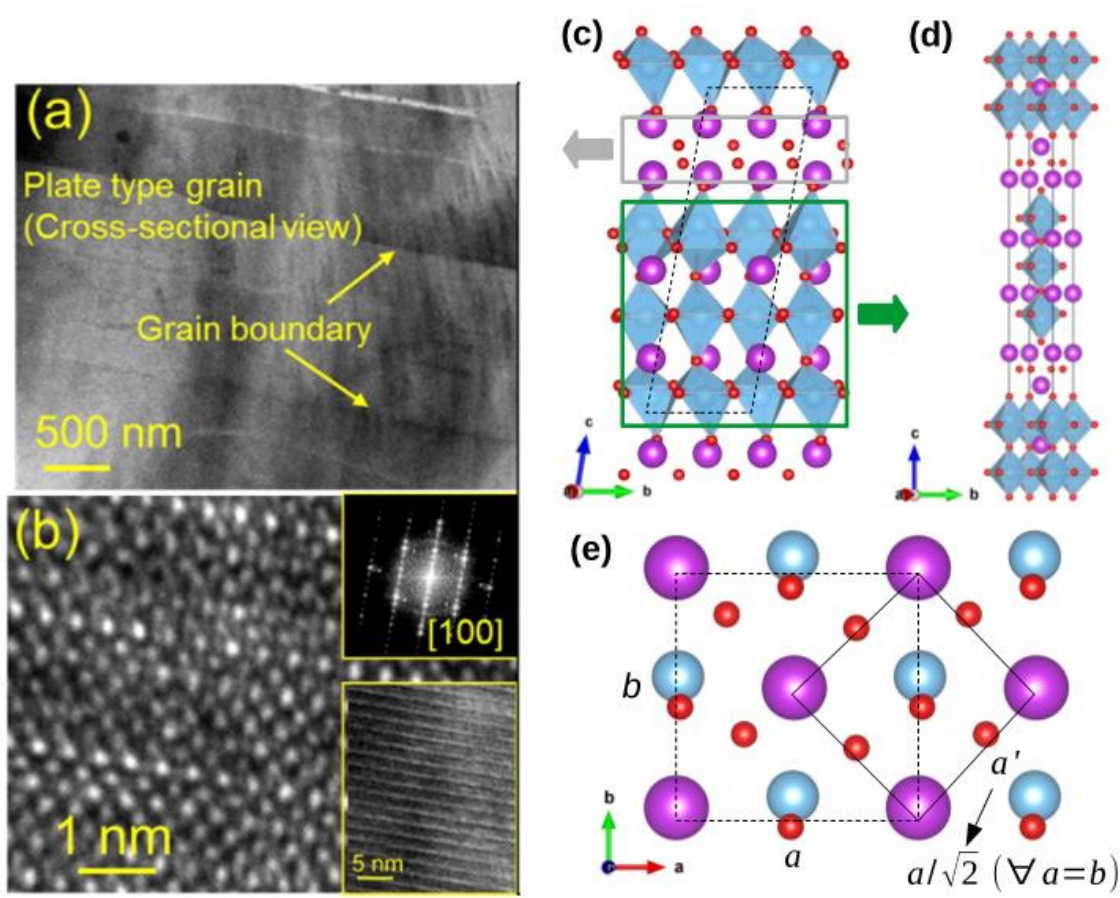

**Figure 1** (a) Bright field cross-section TEM image of plate type grains in BiT indicates that the thickness is in the range of 200–500 nm. (b) The HR-TEM lattice fringe images of BiT ceramics observed from zone axis [100] indicate the stacking of the pseudo-perovskite and $(Bi_2O_2)^{2+}$ layers. The lower inset of (b) shows the corresponding low magnification image. Note that images of $Bi_2O_2$ layers in the HR-TEM image are collected with the electron beam parallel to the [100] zone axis. The upper inset of (b) depicts the corresponding FFT patterns indicating [100] zone axis. Low and high temperature phases of the relaxed BiT structures are shown in (c) and (d), respectively. Bi is denoted by large purple spheres, O by small red spheres. Ti ions stay at the center of the light blue octahedral surrounded by six O atoms. (e) A suggested transformation path from monoclinic to tetragonal symmetries. This transition is associated with the opposite movement of the fluorite- and perovskite-like layers, indicated by gray and green arrows shown in (c), respectively.

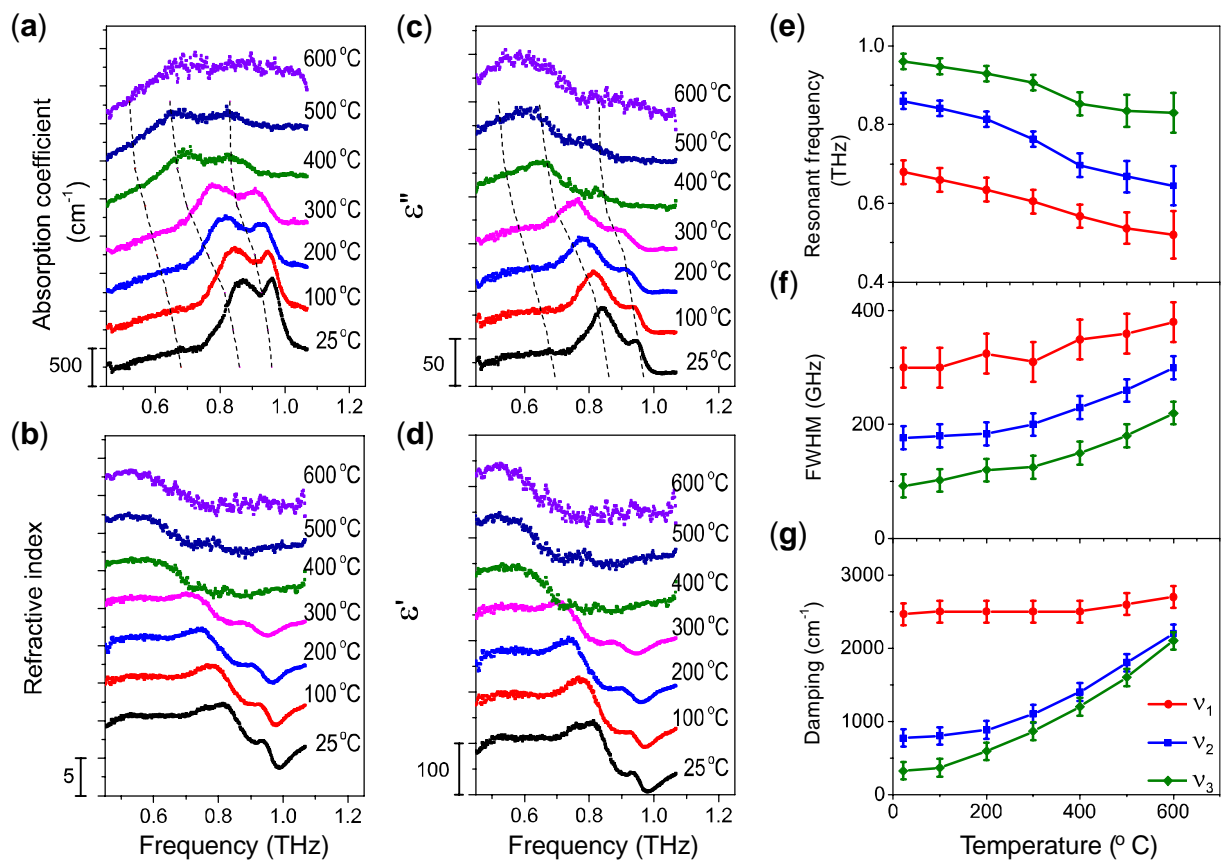

**Figure 2:** The terahertz (a) absorption and (b) refractive index of the *c*-oriented textured polycrystalline BiT ceramic material were recorded at various temperatures. Complex terahertz dielectric response including (c) the dielectric loss and (d) the permittivity at different temperatures calculated from their absorption and refractive index provides insight into the structural dynamics of the BiT material. Employing the three-damped oscillator model, we extracted values for optical phonons (e) soft phonon frequencies $\nu_1$, $\nu_2$, and $\nu_3$, (f) FWHM and (g) phonon damping factors $\gamma_1$, $\gamma_2$, $\gamma_3$. The curves are shifted for clarity in panels (a-d) and the dash lines are guide for eye.

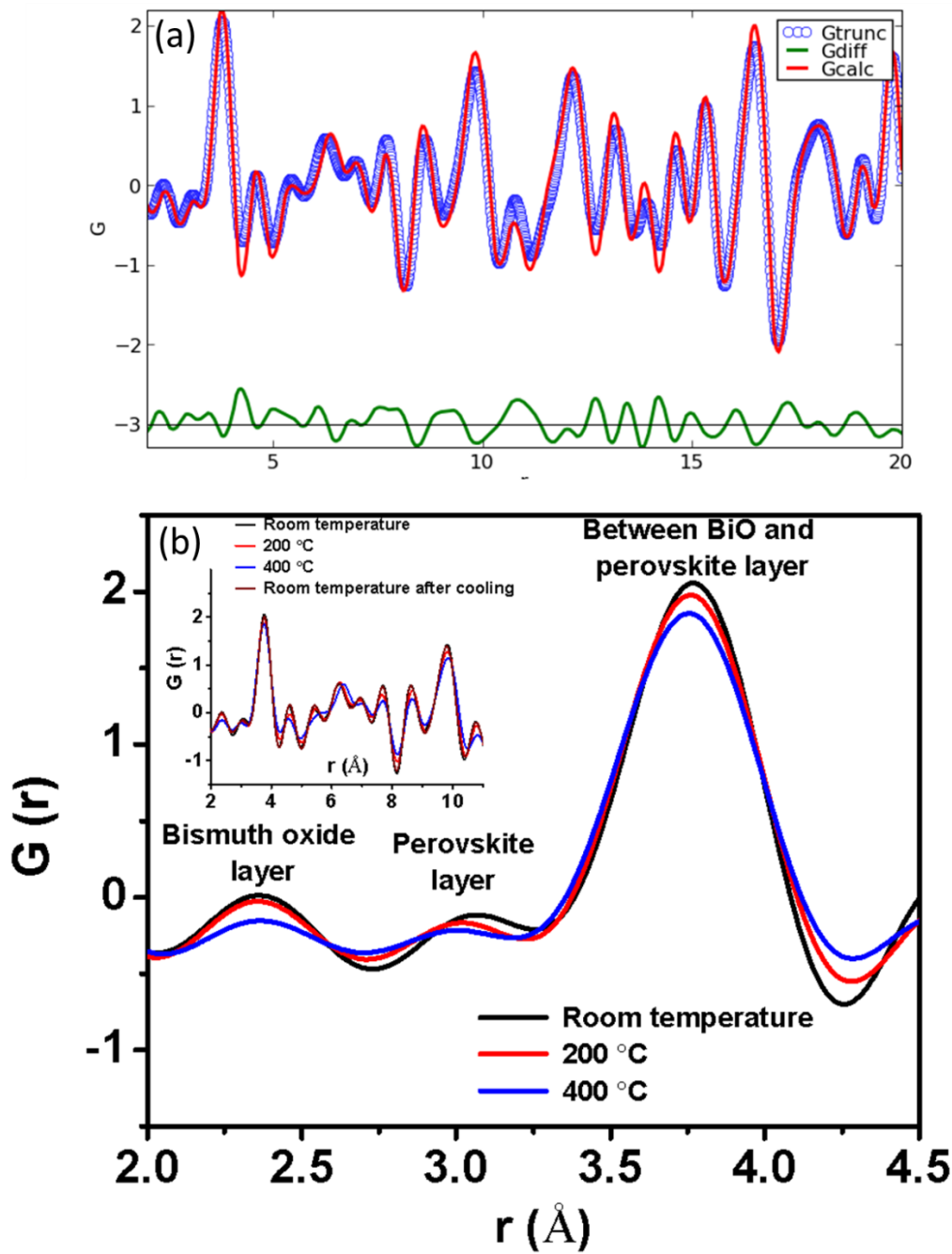

**Figure 3:** (a) The fit obtained using PDFGUI for B2cb structure in BiT. (b) Peaks indicate the closest neighbor Bi−O bonds. The Bi−O bonds show significant disordered structure at higher temperatures for both bismuth oxide and the perovskite layers. The inset of Fig. 3(b) show the pair distribution functions, G(r), measured under different conditions, providing a relation between the dynamics of Bi ions with phonon dynamics. The calculated pattern for the B2cb structure (high temperature orthorhombic phase) is shown with dotted line.

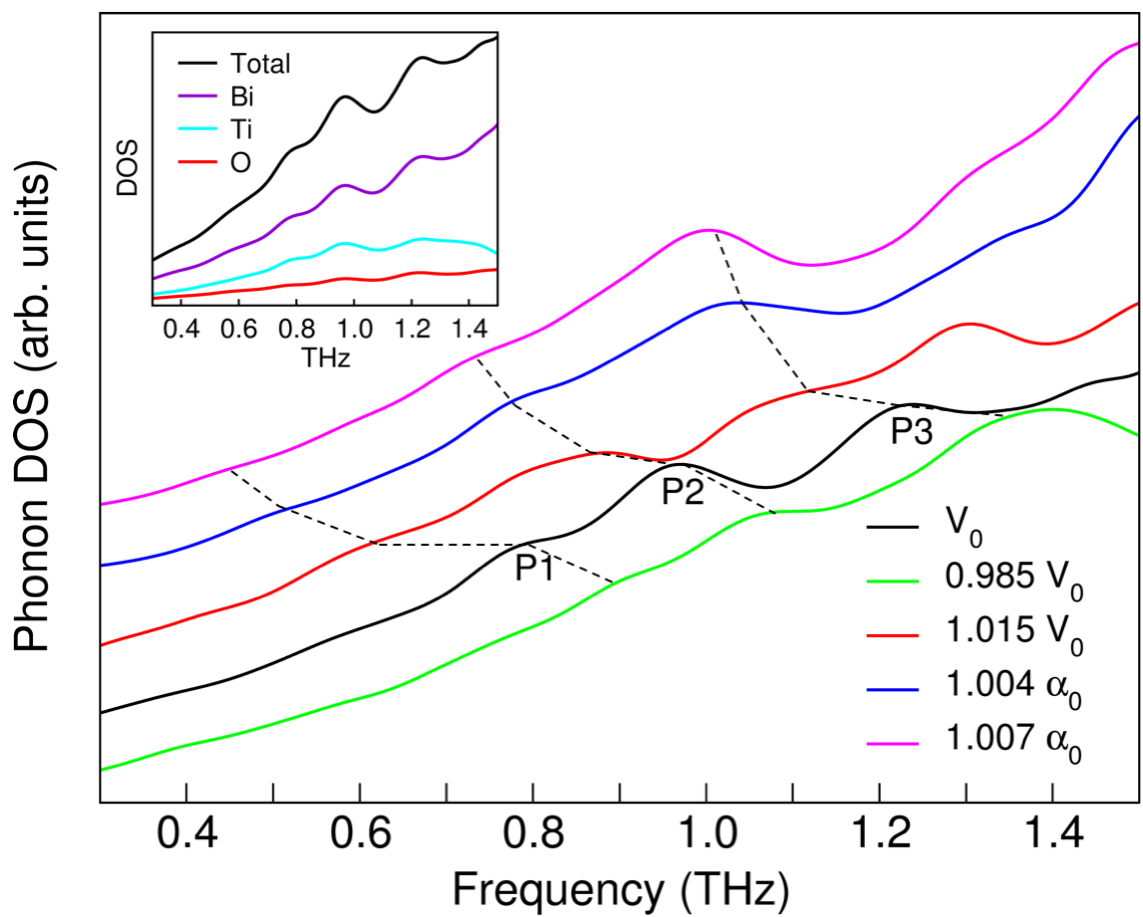

**Figure 4:** Phonon density of states (DOS) calculated using frozen phonon method. The phonon DOS for ground state monoclinic structure ($V_0$) is shown in black solid line. Hydrostatic change in volume by -1.5% ($0.985\ V_0$) and +1.5 ($1.015\ V_0$) are shown as green and red solid lines, respectively. The DOS for the change in monoclinic angle by 0.4% ($1.004\ \alpha_0$) and 0.7% ($1.007\ \alpha_0$) from ground state ($\alpha_0$) are shown in blue and magenta solid lines, respectively. The peaks shift to lower frequencies in all cases due to rearrangement of atomic positions upon relaxation (see SM). The dashed lines are guide to the eye. The inset shows atomic contribution to the total phonon DOS, suggesting that a major contribution to phonons in the low frequency range is due to the Bi atoms. The DOS is shifted for clarity.

**Supplemental material**

# Soft Phonon Mode Dynamics in Aurivillius Type Structures

Deepam Maurya[1], Ali Charkhesht[2], Sanjeev K. Nayak[3], Fu-Chang Sun[3], Deepu George[2], Abhijit Pramanick[4], Min-Gyu Kang[1], Hyun–Cheol Song[1], Marshall M. Alexander[2], Djamila Lou[2], Giti A. Khodaparast[2], S. Pamir Alpay[3], N. Q. Vinh[2], and Shashank Priya[1]

[1]*Bio-inspired Materials and Devices Laboratory (BMDL), Center for Energy Harvesting Materials and Systems (CEHMS), Virginia Tech, Blacksburg, VA 24061 USA*

[2]*Department of Physics, Virginia Tech, Blacksburg, VA 24061 USA*

[3]*Department of Materials Science & Engineering, Department of Physics, Institute of Materials Science, University of Connecticut, Storrs, CT 06269-3136, USA*

[4]*Department of Physics and Materials Science, City University of Hong Kong, Kowloon, Hong Kong SAR*

---

## 1. Sample preparation:

In order to synthesize textured $Bi_4Ti_3O_{12}$ (BiT) ceramics, we first synthesized BiT platelets using molten salt synthesis method [1]. For this, stoichiometric amount of $Bi_2O_3$ and $TiO_2$ ceramics were ball milled for 24 h under ethyl alcohol in polyethylene bottle with yttria-stabilized zirconia (YSZ) balls as milling media. This slurry was further dried in an oven at 60ºC for 6 h. The resulting powder was mixed with equal amount of salt mixture (56wt% KCl and 44wt% NaCl) and heated at 1150 ºC for 30 min. Next, this product is washed several times in deionized water to remove salt and get BiT platelets. These platelets were pressed to get cylindrical pellet and sintered at 1150 ºC for 2 h. Furthermore, during high temperature processing, all the specimens were muffled with the powder of the same composition in order to maintain the chemical composition. During pressing, the anisotropic BiT particles with large aspect ratio were aligned with the major surface perpendicular to the pressing direction, which resulted in textured BiT ceramics after sintering.

## 2. Texture and microstructure

Room temperature XRD-spectra were recorded by using a Philips Xpert Pro x-ray diffractometer (Almelo, The Netherlands). These XRD patterns clearly suggest formation of pure BiT phase. The change in the intensity of Bragg reflections in the XRD-spectrum of textured BiT clearly suggest high degree of texturing. The degree of orientation was determined from the XRD pattern of the textured BiT in the range of $2\theta = 10 - 50^o$ by Lotgering's method. The Lotgering factor $f$ is defined as the fraction of area textured with required crystallographic plane using the formula [2]:

$$\text{Lotgering Factor } f_{00l} = \frac{P - P_o}{1 - P_o}, P = \frac{\Sigma I(00l)}{\Sigma I(hkl)}, P_o = \frac{\Sigma I_o(00l)}{\Sigma I_o(hkl)} \quad (1)$$

where $I$ and $I_o$ are intensity of the diffraction lines ($hkl$) of textured and randomly oriented specimens, respectively. The degree of texturing (calculated using Lotgering factor) was found to be 96% suggesting sintered ceramics are oriented along c-axis.

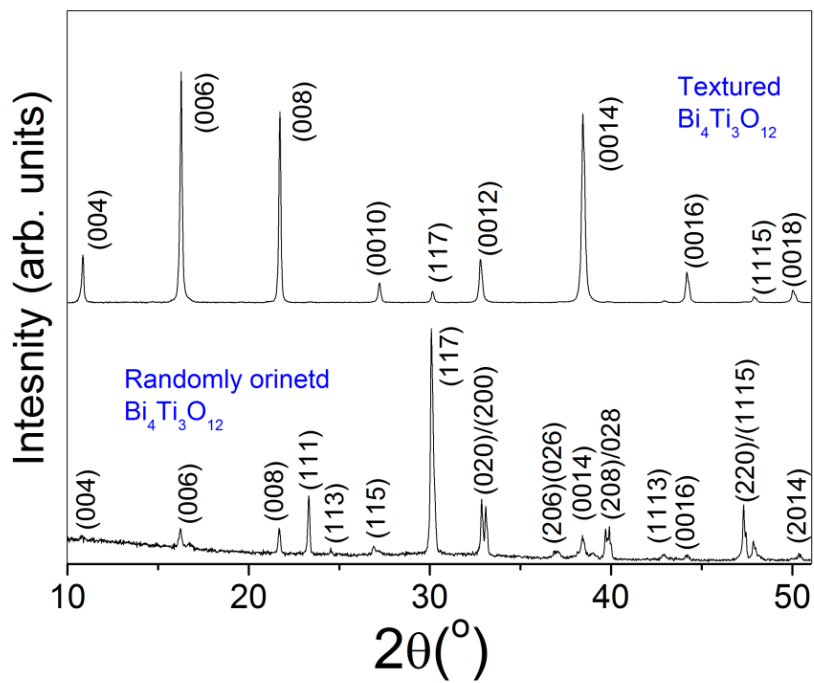

**Figure S-1** XRD spectra recorded at RT for textured and randomly oriented BiT ceramics. Please note the change in the intensity of textured BiT ceramics indicating high degree of the crystallographic orientation along *c*-axis.

The surface morphology of the sintered samples was observed using a LEO Zeiss 1550 (Zeiss, Munich, Germany) scanning electron microscope. In order to prepare the electron transparent TEM specimens, we used standard grinding and ion-milling method. For conducting transmission electron microscopy, we used a FEI Titan 300 microscope.

Figure S-1 shows XRD-patterns recorded at room temperature for textured BiT ceramics. The change in the intensity of various Bragg reflections compared to randomly oriented counterpart, suggests a high degree of texture along $c$-axis. The lotgering factor $f_{00l}$ calculated from these XRD-patterns suggested 96% texturing in <001> orientation. In order to investigate morphology of the textured BiT samples, SEM micrographs were recorded on the flat surface (Fig. S-2$a$). The flat surface clearly shows a plate type (major surface in the plane of the sample's top surface) grain morphology. The cross section morphology indicates that the plate type BiT grains are stacked along the thickness of the sample. This morphology confirms the textured microstructure of BiT. From these images it can be seen that the size of the plate-type grains was in the range of 5-15 μm. The thickness of these plate type grains was found to be in the range of 200-500 nm (Fig. S-2(a)). Figures S-2(b) and S-2(c) show the bright field cross section TEM micrographs of a plane view samples. The grain boundaries in textured BiT are the result of fusing BiT plates together (during high temperature sintering process) mostly with a slight miss-orientation. Fig. S-2(d) show a HR-TEM image of the lattice fringes from [001] orientation. Due to the twofold in-plane symmetry, the distinctive stacking of the pseudoperovskite and $(Bi_2O_2)^{2+}$ layers was not observed. Moreover, the layered structure of these materials was found to be useful in decreasing the thermal conductivity due to effective phonon scattering.[3]

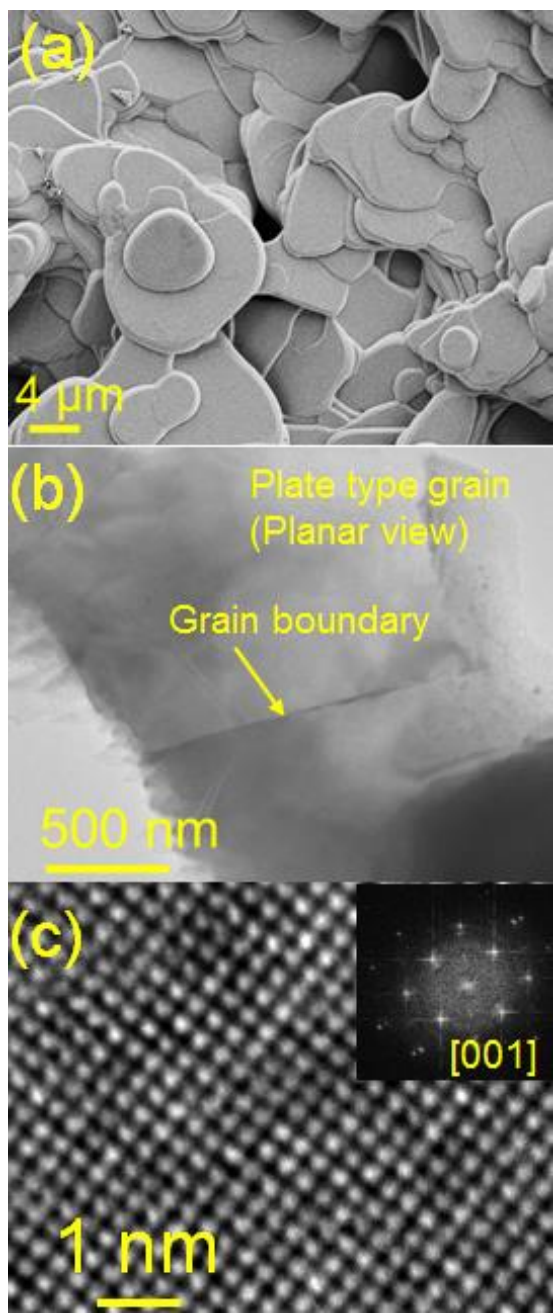

**Figure S-2** (a) SEM micrographs of <001> textured BiT from top surface. (b) Bright field TEM image of the plate type grains from the top surface. (c) The HR-TEM lattice fringe images of BiT ceramics from [001] zone axis. The inset of (c) shows FFT patterns marked with the zone axes.

## 3. Terahertz (THz) measurements:

The details of the THz spectrometer measurements can be found elsewhere [4]. The spectrometer supports the simultaneous measurements of absorbance and refractive index of solutions over the spectral range from 26.5 GHz to 1.12 THz (0.88 to 37.36 cm$^{-1}$ or 0.268 to 11.3 mm). The signal-to-noise and spectral resolution of this device were significantly high as compared to any previous state-of-the-art instruments. For example, while the dynamic range of a commercial terahertz time-domain spectrometer is just $10^6$ and its spectral resolution is several gigahertz, the dynamic range of our instrument reaches an unprecedented value of $10^{13}$ and the system achieves a spectral resolution of less than 100 Hz [4-6]. The system provides a coherent radiation source with a power up to 20 mW in the gigahertz-to-terahertz region.

For transmission measurements, we employed a quasi-optical setup. The transmitter module emits terahertz radiation into free space with a circular horn. The radiation transmitted through the sample is subsequently collected using a similar horn and is fed into a receiver module. For temperature dependent measurements, we used a home built temperature controllable setup made of a large aluminum block and high power resistances. Temperature of the sample was controlled with an accuracy of $\pm 2$ $^o$C by varying the voltage applied to the power resistances. The BiT film was placed on a brass sample holder using thermal paste and was attached to an aperture on the setup. An identical aperture and empty sample holder formed reference signals. The translation stage allowed alternative reference and sample measurements under identical conditions. The high dynamic range combined with the ability to detect the phase allowed an accurate measurements of absorption coefficient and refractive index of highly absorptive samples like BiT. The experiment consists of two consecutive measurements for each temperature: (i) Measurements of the reference signals of transmitted intensity and phase shift with an empty sample holder. (ii) Measurements of intensity and phase shift after the BiT sample.

## 4. Atomic Pair Distribution Function (PDF) results:

The atomic pair distribution function analysis can be used to understand local structural changes. PDFgetX was used to compute G(r) from the X-ray diffraction data recorded on $Bi_4Ti_3O_{12}$, as shown in Fig. S-3. The $Q_{maxinst}$ was limited to 10 $Å^{-1}$ due to the increased noise. The wavelength used for calculations is 0.414211 Å.

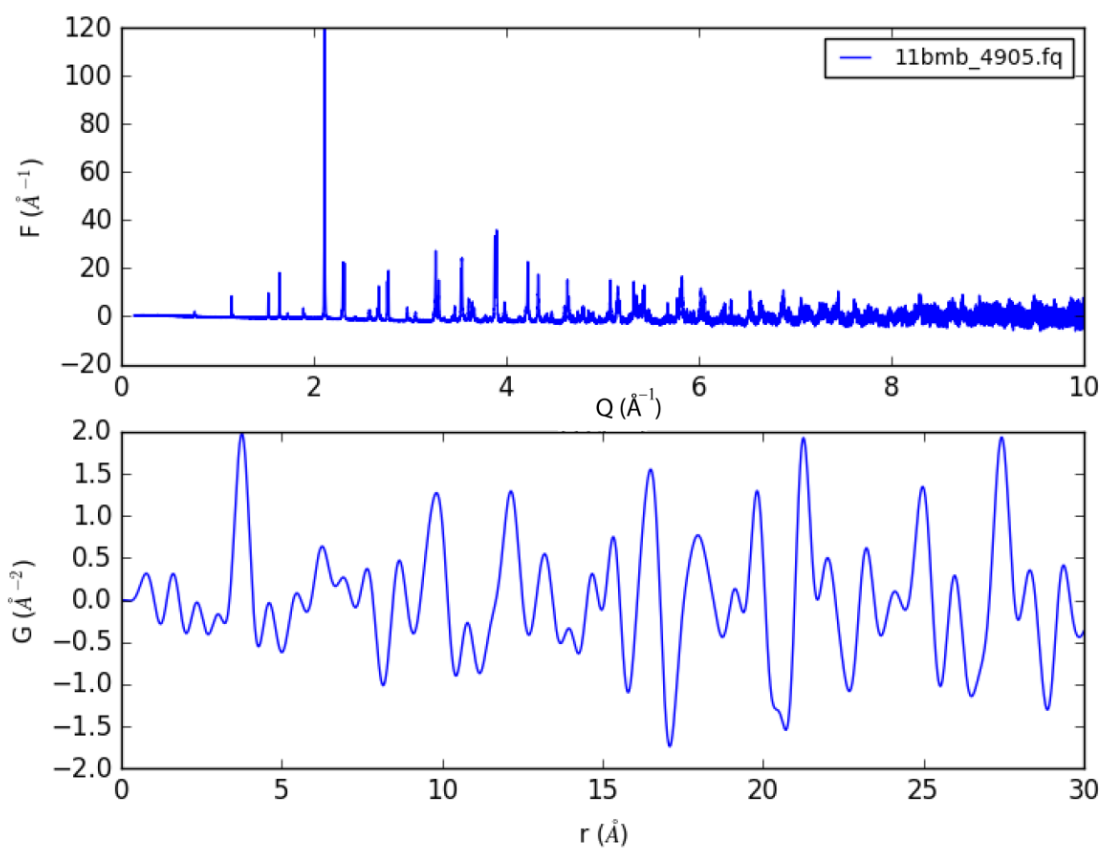

**Figure S-3**: F(Q) and G(r) profiles for BiT from PDFgetX [7].

## 5. Raman measurements:

Raman measurements were conducted at the Virginia Tech Vibrational Spectroscopy Laboratory using a Jobin-Yvon HR800 Raman microprobe equipped with a 514 nm laser focused onto the sample through a microscope lens. Machado *et al*. [8] reported two unstable $E_u$ modes in BiT, one involving the motion of $(Bi_2O_2)^{2+}$ layers in a direction relative to the perovskite blocks and the second one involving the motion of the Bi ions in the perovskite at A site with respect to the $TiO_6$ perovskite blocks. The unstable $E_u$ mode in layered materials involving movements of $(Bi_2O_2)^{2+}$ layers with respect to the perovskite block, is a general feature of the Aurivillius compounds [8]. These modes occur at very low frequency zone of center optical phonons, in which layers move as rigid units. The remaining unstable mode in BiT has, however, completely different displacement pattern. In order to understand phonon dynamics and the local structural changes with temperature, the temperature dependent Raman spectra (Fig. S-4) were recorded.

The heavy Bi ions are expected to exhibit their contribution in the lower frequency regime. The softening of modes at 117 cm$^{-1}$ and 271 cm$^{-1}$ can be treated as a signature that the system undergoes structural transitions. Modes above 200 cm$^{-1}$ stem from $TiO_6$ octahedra. The Raman modes at 850, 617, 570, 330 cm$^{-1}$ and 450, 354 cm$^{-1}$ are assigned to $A_{1g}$ and $B_{1g}$ modes, respectively. The modes at 537, 271, 225 cm$^{-1}$ are assigned to $B_{2g} + B_{3g}$ modes originating from the lifting of $E_g$ degeneracies. The disappearance and changes in the modes at 617, 570, 450, 330, 225, and 183 cm$^{-1}$ are believed to be due to the reconstructions in overcoming the distortion and octahedral tilting. These changes hint the nanoscale structural changes in the system, which could be the precursors for the structural transformation at the Curie temperature.

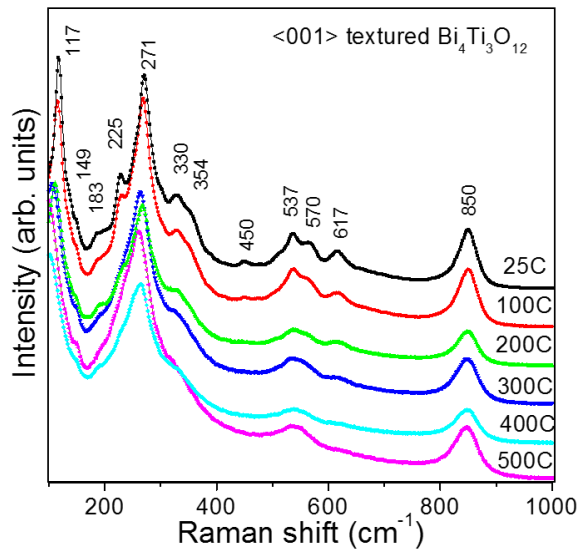

**Figure S-4** Raman spectra recorded as function of temperature. Change in frequency and intensity of various modes reflects structural changes with respect to the temperature.

## 6. Density functional theory based calculations

We carried out first-principles calculations using density functional theory (DFT) as implemented in the Vienna *ab initio* Simulation Package (VASP). The generalized gradient approximation (GGA) parameterized by Perdew-Burke-Ernzerhof was used for the exchange correlation functional together with the projector augmented wave method. The **k**-point mesh was taken as 8×8×2 with Γ centered Monkhorst-Pack grids, along with the plane wave energy cutoff 500 eV. The energy tolerance for self-consistent field calculations was set to $10^{-6}$ eV. The lattice parameters were found to be $a = 5.49$ Å, $b = 5.53$ Å, $c = 16.88$ Å, and $\alpha = 80.61°$ (referred as $\alpha_0$ in main text to refer to ground state monoclinic structure) for the minimum energy crystal structure in space group *P*c. Phonon calculations were done by two main approaches. The first approach is the frozen phonon method, where the lattice was expanded to 2×2×1 supercell and the displacement vectors were created for the monoclinic space group using the Phonopy software. The forces for each displacement configuration was calculated using VASP and fed back to Phonopy in order to construct the dynamical matrix and compute the band structure and density of states (DOS). The DOS was calculated by integration over the Brillouin zone with a 30×30×30 k-points grid and a smearing factor of 0.04. In a second approach, the phonon eigenvalues were calculated using the density functional perturbation theory (DFPT) as implemented in VASP. The phonon modes obtained has been used to study the energy as a function of amplitude for the low energy phonon modes.

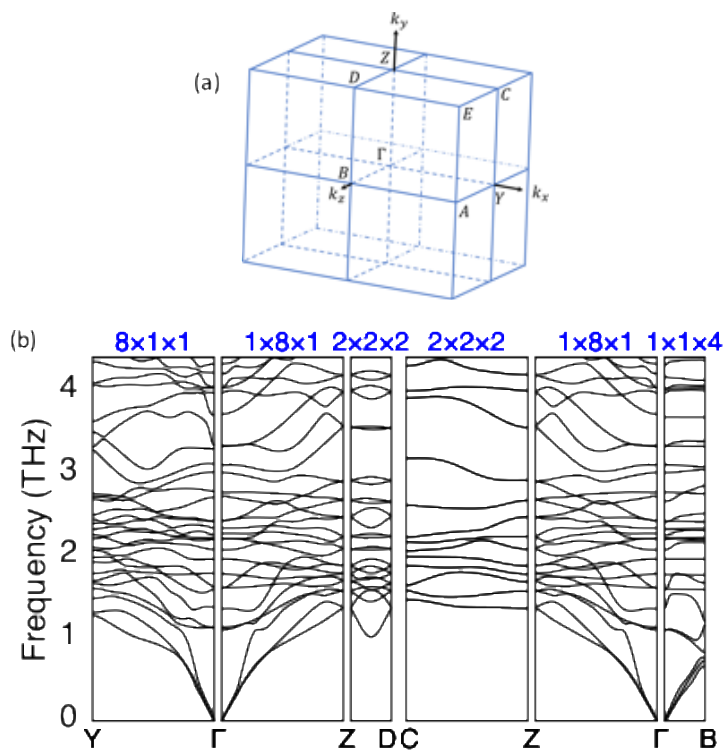

**Figure S-5**: Brillouin zone (a) and phonon band structure (b) for the monoclinic BIT lattice. The appropriate size of supercell is chosen as shown in the top panel to incorporate the commensurate points along the path between high symmetry points.

The phonon band diagram was established by appropriately expanding the size of the supercell to account for all the commensurate points along $x$- ($\Gamma$-Y), $y$- ($\Gamma$-Z), and $z$-axes ($\Gamma$-B), as shown in Fig. S-5. This was found to be in the same direction of lattice parameters, $a$, $b$, and $c$, respectively.

To determine the complete band diagram, computationally feasible supercells of size up to 8×8×4 consisting of 9728 atoms are required which is not computational feasible.

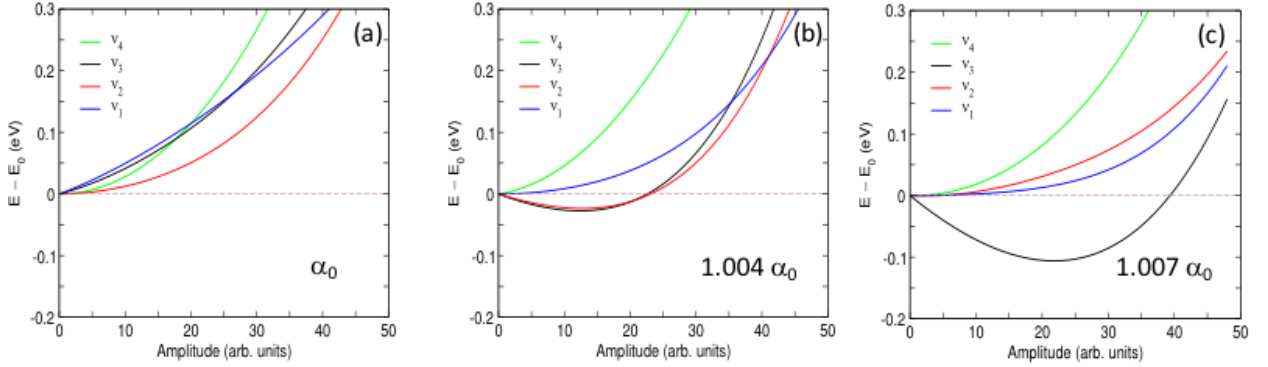

**Figure S-6**: Total energy as a function of amplitude for the lowest four phonon modes for (a) perfect monoclinic structure, (b) monoclinic lattice with α = 1.004 α₀ and (c) monoclinic lattice with α = 1.007 α₀, respectively. The lowering in energy as a function of amplitude is a signature of soft phonon mode.

In order to study the soft phonon characteristics, the modes obtained for the lowest four phonon energies from DFPT calculations have been subject to variation of amplitude and the total energy of the supercells calculated using VASP. The models with pure monoclinic lattice ($\alpha_0$), and lattice with $\alpha = 1.004\ \alpha_0$ and $\alpha = 1.007\ \alpha_0$ were chosen as these models were adapting from monoclininc towards the tetragonal structure, as shown in Fig. S-6. The symmetry of the three modes at the $\Gamma$ point using Quantum Espresso code (see Acknowledgement in main text) is found to be $A''$, $A'$, and $A'$, all of which are IR active. The crystallographic volume change with respect to the change in temperature is depicted in Fig. S-7. The volume change is positive up to certain temperature and then start decreasing with increasing temperature. This is in line with the larger crystallographic volume of the ferroelectric monoclinic phase as compared to the paraelectric tetragonal phase of BiT as the monoclinic phase of BiT possesses larger volume (252.13 Å³) than that of the tetragonal phase (246.34 Å³) of BiT [8].

**Table S-1**: Table showing the lattice parameters of various models considered in this study. Here, $\alpha_0$ and $V_0$ (for two f.u.) are the monoclinic lattice angle and volume for the ground state crystal structure.

| | Volume | a | b | c | $\alpha$ | $\beta$ | $\gamma$ |
|---|---|---|---|---|---|---|---|
| **DFT (present)** | 506.252 | 5.493 | 5.533 | 16.883 | 80.609 | 90 | 90 |
| **1.004 $\alpha_0$** | 506.252 | 5.450 | 5.522 | 17.034 | 80.933 | 90 | 90 |
| **1.007 $\alpha_0$** | 506.252 | 5.436 | 5.525 | 17.060 | 81.148 | 90 | 90 |
| **1.015 $V_0$** | 513.884 | 5.521 | 5.560 | 16.968 | 80.609 | 90 | 90 |
| **1.03 $V_0$** | 521.592 | 5.548 | 5.588 | 17.052 | 80.609 | 90 | 90 |
| **Tetragonal** | 492.676 | 3.852 | 3.852 | 33.197 | 90 | 90 | 90 |
| **Ref. [8]** | 486.652 | 3.85 | 3.85 | 32.832 | 90 | 90 | 90 |

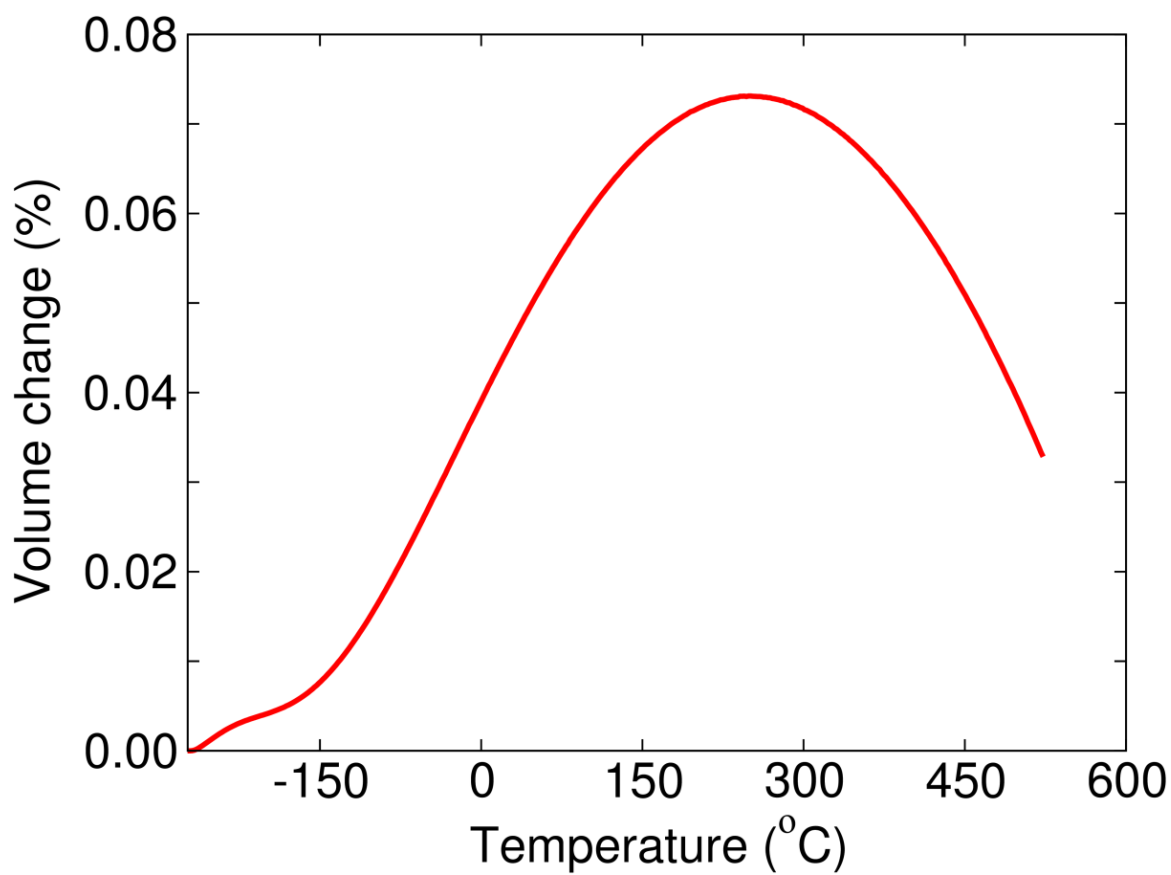

**Figure S-7** Temperature dependence of the crystallographic volume change for BiT obtained from quasi-harmonic approximation.

The Screenshot of the animated gif files showing the phonon modes for the lowest three frequencies is depicted in Fig. S-8.

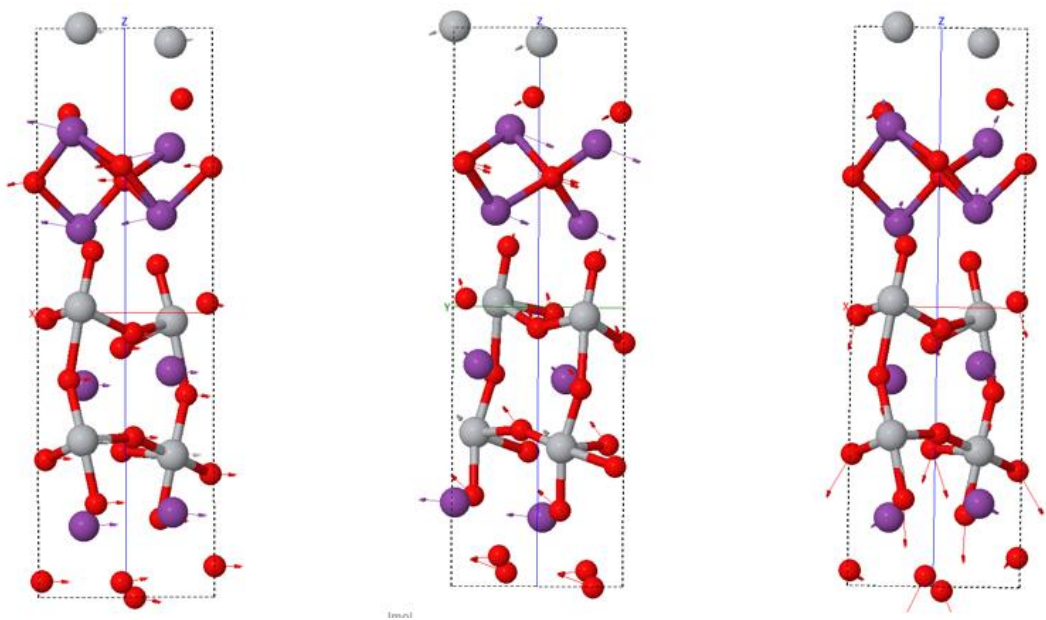

**Fig. S-8** Screenshot of the animated gif files of the lowest frequency vibration modes depicting the phonon modes for the lowest three frequencies.

**References:**

[1]     D. Maurya, Y. Zhou, Y. Yan, and S. Priya, Journal of Materials Chemistry C **1**, 2102 (2013).

[2]     F. K. Lotgering, J Inorg Nucl Chem **9**, 113 (1959).

[3]     C. Chiritescu, D. G. Cahill, N. Nguyen, D. Johnson, A. Bodapati, P. Keblinski, and P. Zschack, Science **315**, 351 (2007).

[4]     D. K. George, A. Charkhesht, and V. N. Q., Review of Scientific Instruments **85, in print** (2015).

[5]     N. Q. Vinh, S. J. Allen, and K. W. Plaxco, J Am Chem Soc **133**, 8942 (2011).

[6]     N. Q. Vinh, M. S. Sherwin, S. J. Allen, D. K. George, A. J. Rahmani, and K. W. Plaxco, J Chem Phys **142** (2015).

[7]     A. D. Rae, J. G. Thompson, R. L. Withers, and A. C. Willis, Acta Crystallographica Section B **46**, 474 (1990).

[8]     R. Machado, M. G. Stachiotti, R. L. Migoni, and A. H. Tera, Phys Rev B **70**, 214112 (2004).